\documentclass[prb,twocolumn,superscriptaddress,notitlepage]{revtex4-2}
\usepackage{CJK}
\usepackage[pdftex]{graphicx}
\usepackage{float}
\usepackage{amssymb}
\usepackage{url}
\usepackage{natbib}
\usepackage{amsmath}  
\usepackage{amsfonts} 
 
\usepackage[colorlinks,
                   linkcolor=blue,
                   anchorcolor=blue,
                   citecolor=blue,
                   urlcolor=blue]
                  {hyperref}      

\begin{document}

\title{Spin Resonance in Perspective of Floquet Theory and Brillouin-Wigner Perturbation Method}

\author{Mingjun Feng}
\affiliation{National Time Service Center, Chinese Academy of Sciences, Xi'an, 710600, China.}
\affiliation{University of Chinese Academy of Sciences, Beijing, 100049, China}

\author{Guobin Liu}
\email{liuguobin@ntsc.ac.cn}
\affiliation{National Time Service Center, Chinese Academy of Sciences, Xi'an, 710600, China.}
\affiliation{University of Chinese Academy of Sciences, Beijing, 100049, China}
\affiliation{Key Laboratory of Time Reference and Applications, Chinese Academy of Sciences, Xi’an 710600, China}

\date{\today}

\begin{abstract}
We studied the two-level spin resonance in a new perspective. Using the Floquet theory, the periodic interaction Hamiltonians were transfromed into a time-independent interaction. Using the Brillouin-Wigner perturbation method, a degenerated subspace is constructed, where the effective Hamiltonian is given in a perturbation expansion. In this framework, we found that the upper triangular element $\langle \alpha | H^1 | \beta \rangle$, determines whether the resonance happens. The generalized Rabi frequency and the Bloch-Siegert shift were solved straightforwardly as the first order and the second order solution, proving the benefit of the developed method.
\end{abstract}

\maketitle 

\section{Introduction} 

Magnetic resonance, or spin resonance is a generally accomodated concept in many discplines of physical science, such as nuclear magnetic resonance, electron paramagnetic resonance and ferromagnetic resonance, etc. It is widely used in subjects from magnetic resonance spectroscopy \cite{Karlsson2022} to quantum information science \cite{Jones2024, Koppens2006, John2024}. The simplest paradigm is a two-level spin interacting with a static magnetic field and an alternating electromagnetic (EM) field. When the energy of the EM field matches the energy gap of the spin particle in the static field, absorption or emission of photons occurs. In a dynamical perspective, the particle undergo transitions between spin up and spin down states continusouly, indicating the occurrence of spin resonance.

The Rabi model \cite{Lahiri2016, Braak2016} is a semiclassical model of spin resonance.
It can usually be solved through the well known method of rotating wave approximation (RWA) \cite{Lahiri2016, Jaynes1963}.
In RWA, the fast oscillating term, namely the counter-rotating term of the Hamiltonian, is omitted under the condition of weak spin-field interaction. Unlike the rotating wave approximation, which ignores high-frequency oscillations, our method explains from another perspective why the counter-rotating term can be ignored.

Floquet theory \cite{Shirley1965} is also used to solve the Rabi model.
It can transform the Schr\"{o}dinger equation with the periodic Hamiltonian into a time-independent Schr\"{o}dinger equation.
Then, time-independent perturbation method is used to further solve the equation.
Unlike existing methods \cite{Shirley1965} using Salwen's perturbation method \cite{Salwen1955}, we instead use a more general perturbation method \cite{Hubac2010}, which has a concise form, making the calculation process simple and clear.
Besides calculating the Rabi model, we also calculate the transition probabilities of spin particles in three other electromagnetic wave modes.
By analyzing the first-order perturbation results, we find that the occurrence of resonance depends on 
the upper triangular element $\langle \alpha | H^1 | \beta \rangle$, determined by the first positive Fourier component of the periodic Hamiltonian.
If the element is zero, it will result in resonance occurring, while a non-zero element leads to the absence of resonance.
Finally, by calculating the second-order transition probability, we get the Bloch-Siegert shift \cite{Bloch1940} of spin resonance for the linearly polarized field.
Students and researchers may gain some new understanding of spin resonance from our computational method.

\section{Model} \label{Model}
The time-dependent Schr\"{o}dinger equation is
\begin{equation}
i \hbar \frac{d}{dt} |\psi(t)\rangle = \mathcal H(t)  |\psi(t)\rangle
\label{SE}
\end{equation}

For the Rabi model, the time-dependent magnetic dipole interaction Hamiltonian $H(t)$ can be expressed by
\begin{equation}
\mathcal H (t)  =  - \boldsymbol \mu\cdot \mathbf B=-\gamma \mathbf S\cdot \mathbf B=-\gamma\frac{\hbar}{2}  \boldsymbol \sigma\cdot \mathbf B
\end{equation}
where $\boldsymbol \mu$ denotes the magnetic moment, $\gamma$ stands for the magnetogyric ratio, $\mathbf S$ represents the spin 
operator, $\boldsymbol \sigma = (\sigma_x, \sigma_y, \sigma_z) $ signifies the Pauli spin operator.

In typical spin resonance experiments, the static magnetic field is set in the longitudinal direction, i.e. the $z$ direction, while the alternating EM field is set in a transvere direction, i.e. the $x$ or $y$ direction, leading to the most common form of the magnetic field vector is $\mathbf B =  (B_1 \cos \omega t, 0, B_0)$, so the first kind of spin resonance interaction Hamiltonian is written as
\begin{eqnarray}
\mathcal H_{1}(t) & = & - \gamma \frac{\hbar}{2}(B_0\sigma_z+B_1\sigma_x\cos \omega t),
\end{eqnarray}

For simplification, we introduce the intrinsic spin resonance frequency $\omega_0=-\hbar \gamma B_0$ and the spin-field coupling strength $b=-\hbar \gamma B_1/4$, thus we get
\begin{equation}
\mathcal H_{1}(t)=\frac{1}{2} \omega_0\sigma_z + 2b \sigma_x\cos \omega t.
\label{H1}
\end{equation}
We assume that the magnetogyric ratio is negative, corresponding to $b>0, \omega_0>0$, and finial outcomes of a positive magnetogyric ratio will be similar.

\begin{figure}[h!]
\centering
\includegraphics{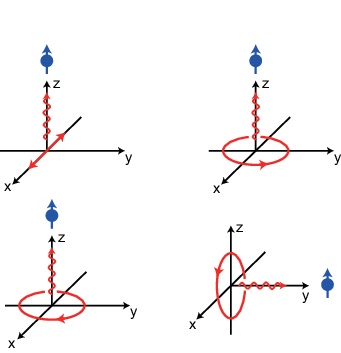}
\caption{The coupling of spin particles to four electromagnetic wave modes.
The red lines denote the direction of the wavevector and the polarization type of electromagnetic waves.
The static magnetic field along the $z$-axis is omitted in the figure.
}
\label{fig1}
\end{figure}

We also consider the other three types of Hamiltonians,
\begin{eqnarray}
\mathcal H_{2}(t) & = & \frac{1}{2} \omega_0\sigma_z + b (\sigma_x\cos \omega t + \sigma_y\sin \omega t) ;
\label{H2}
\\
\mathcal H_{3}(t) & = & \frac{1}{2} \omega_0\sigma_z + b (\sigma_x\cos \omega t - \sigma_y\sin \omega t) ;
\label{H3}
\\
\mathcal H_{4}(t) & = & \frac{1}{2} \omega_0\sigma_z + 2b (\sigma_x\cos \omega t + \sigma_z\sin \omega t).
\label{H4}
\end{eqnarray}
These four Hamiltonians in Eqs.(\ref{H1}-\ref{H4}) describe the interaction between spin particles and four modes of electromagnetic waves, as shown in Fig.\ref{fig1}.
Classically, the concept of resonance is understood as a small drive leading to a large response when the driving frequency approaches the intrinsic frequency of the driven system.Therefore, like in the RWA, here we also focus on weak driving situations in our model, saying the driving strength is far smaller than the intrinsic frequency, i.e. $b \ll \omega_0$. 

\section{Floquet Theory} \label{Floquet}
As the Hamiltonians $\mathcal H(t)$ in Eqs.~(\ref{H1}-\ref{H4}) is periodic, it is convenient to solve the Schr\"{o}dinger equation Eq.~(\ref{SE}) using the Floquet theory \cite{Shirley1965}.
The problem will be equivalently transformed as determining the eigenvalues and eigenvectors of a time-independent Hamiltoninan.
Although the dimension of Floquet matrix is infinite, the arrangement of its matrix elements is regular, allowing the problem to be further solved.

In the framework of Floquet theory, the periodic Hamiltonian needs to be written as a Fourier series expansion.
In our model, the Fourier series includes only three terms,
\begin{equation}
\mathcal H(t)=H^0 + H^1e^{i\omega t} + H^{-1}e^{-i\omega t},
\label{eq:periodicHamiltonian}
\end{equation}
where the first term represents the intrinsic Hamiltoninan, the second and third terms represent the positive and negative Fourier components of the interaction Hamiltonian, whose complex amplitudes have the relation $H^{-1} = ( H^1 )^\dagger$.

Based on Fourier coefficients, we can write the Floquet matrix $H$.
Especially, our model is the perturbation problem, so we can write the Floquet matrix as two terms
\begin{equation}\label{eq:FloquetHamiltonian}
H  = H_{0} +V,
\end{equation}
where the non-perturbative Hamiltonian is 
\begin{eqnarray}
H_{0}= 
\begin{pmatrix}
\cdot & \cdot & \cdot & \cdot &  &  & \\
\cdot & H^0-2\omega I & 0 & 0 & \cdot &  & \\
\cdot & 0  & H^0-\omega I &0 & 0 & \cdot & \\
\cdot & 0 & 0 & H^0 & 0 & 0 & \cdot\\
  &  \cdot & 0 & 0 & H^0+\omega I & 0 & \cdot\\
  &  &   \cdot& 0 & 0 & H^0+2\omega I &\cdot \\
  &  &  & \cdot & \cdot & \cdot & \cdot
\end{pmatrix},
\end{eqnarray}
with
\begin{eqnarray}
H^0=
\begin{pmatrix}
E_\alpha & 0
\\
0 & E_\beta 
\end{pmatrix}=
\begin{pmatrix}
\frac 1 2 \omega_0 & 0
\\
0 & -\frac 1 2 \omega_0
\end{pmatrix},
\end{eqnarray}
and $I$ the identity matrix. It is easy to notice that $H_0$ is a diagonal matrix, with eigenvalues $E_{\alpha, n} = E_\alpha + n\omega, E_{\beta, n}=E_\beta+n\omega$ and the corresponding eigenvectors can be denoted as $\{  |\alpha, n\rangle, |\beta, n\rangle  \}$, constituting a complete set. The states $\{  |\alpha, n\rangle, |\beta, n\rangle  \}$ is called as Floquet states.
These states can be explained physically by considering the electromagnetic waves in the model as a quantized field \cite{Shirley1965}.
The integer $n$ represents the number of photons, and $\alpha, \beta$ denote spin states.
Therefore, although our model considers semi-classical Hamiltonians, the Floquet theory allows us to obtain fully quantum results.
We shall only consider the transition process involving the emission or absorption of a single photon.
A particle in a high-energy state may emit a photon, jumping from state $|\alpha, N\rangle$ to state $|\beta, N+1\rangle$, while an atom in a low-energy state may absorb a photon, transitioning from state $|\beta, N+1\rangle$ to state $|\alpha, N\rangle$.
Multiple-photon processes may occur, such as transition from state $|\alpha, N\rangle$ to state $|\beta, N+2\rangle$, whose probability can be proven to be very small in weak driving situations, and are thus ignored directly here.

The perturbative Hamiltonian is
\begin{eqnarray}
V = 
\begin{pmatrix}
\cdot & \cdot & \cdot & \cdot &  &  & \\
\cdot & 0 & H^{-1} & 0 & \cdot &  & \\
\cdot & H^1  & 0 & H^{-1} & 0 & \cdot & \\
\cdot & 0 & H^1 & 0 & H^{-1} & 0 & \cdot\\
  &  \cdot & 0 & H^1 & 0 & H^{-1} & \cdot\\
  &  &   \cdot& 0 & H^1 & 0 &\cdot \\
  &  &  & \cdot & \cdot & \cdot & \cdot
\end{pmatrix}.
\end{eqnarray}

For Floquet Hamiltonian $H$, we can denote its eigenvalues by $\lambda_{\gamma l}$ and eigenvectors by $|\lambda_{\gamma l}\rangle$.
The corresponding time evolution operator is \cite{Shirley1965}
\begin{equation}\label{eq:TimeEvolutionOperator}
U_{\beta \alpha}(t;t_{0})=\sum_{\gamma l} \langle\beta 1|\lambda_{\gamma l}\rangle  \exp[-i\lambda_{\gamma l}(t-t_{0})] \langle\lambda_{\gamma l}|\alpha0\rangle e^{i\omega t},
\end{equation}
which represents the amplitude that a system in state $|\alpha, N\rangle$ at time $t_0$ evolve to state $|\beta, N+1\rangle$ at time $t$.
The transition probability is
\begin{eqnarray}\label{eq:TransitionProbability}
P_{\alpha \to \beta}(t;t_{0})=| U_{\beta \alpha} (t;t_{0}) |^{2}
\end{eqnarray}

In the following section, perturbation method will be employed to determine components $\langle\beta 1|\lambda_{\gamma l}\rangle,  \langle\lambda_{\gamma l}|\alpha0\rangle$ and eigenvalues $\lambda_{\gamma l}$ in Eq.~(\ref{eq:TimeEvolutionOperator}).

\section{Brillouin–Wigner Perturbation Method}
In this section, we use the Brillouin–Wigner perturbation method \cite{Hubac2010, Chen2021} to get eigenvalues and eigenvectors of Floquet Hamiltonian $H$.
The perturbation method uses the so-called partitioning technique, developed independently by Feshbach \cite{Feshbach1958,Feshbach1958a,Feshbach1962} and L\"{o}wdin \cite{Loewdin1965}.
This technique uses projection operators to divide the wave function into two parts, in a model space and an orthogonal space respectively.

We consider $\omega$ is near $\omega_0$, and set the near resonance condition
\begin{equation}
\omega_0 - b < \omega <  \omega_0 + b
\end{equation}
It represents conservation of energy, with $E_\alpha \approx E_\beta+\omega$.
Under the condition, the two unperturbed states $|\alpha, 0\rangle, |\beta, 1\rangle$ are  almost degenerate.

We define projection operator as
\begin{equation}\label{eq:ProjectionOperator}
P = |\alpha, 0\rangle \langle\alpha,0| +  |\beta, 1\rangle \langle\beta,1|, 
\end{equation}
and its orthogonal complement operator
\begin{equation}
P^\perp = \sum_{n\neq 0} |\alpha, n\rangle \langle\alpha,n| + \sum_{n\neq 1}  |\beta, n\rangle \langle\beta,n| .
\end{equation}
The projection operator $P$ will project the state $|\psi\rangle$ onto the degenerate subspace $\mathcal{P}$ with $|\alpha, 0\rangle, |\beta, 1\rangle$ as the basis vectors.

Based on the partitioning technique, the eigenvalues equation in subspace $\mathcal{P}$ is \cite{Hubac2010, Chen2021}
\begin{equation}
H_\text{eff}( \lambda ) (P | \lambda \rangle) = E(P| \lambda \rangle)	
\end{equation}
where $\lambda, | \lambda \rangle$ are the eigenvalue and eigenvector of Floquet Hamiltonian $H$.
The effective Hamiltonian can be written as
\begin{eqnarray}
H_{\mathrm{eff}}(\lambda) & = & P H P +P H P^\perp\frac1{\lambda-P^\perp H P^\perp} P^\perp H P
\nonumber\\
 & = &H_{PP} + H_{PP^\perp} \frac1{\lambda- H_{P^\perp P^\perp} }  H_{P^\perp P }.
\end{eqnarray}
Here we abbreviate $P H P$ as $H_{PP}$, and similarly for other operators.
Using operator identity 
\begin{equation}
\left(X-Y\right)^{-1}=X^{-1}+X^{-1}YX^{-1}+X^{-1}YX^{-1}YX^{-1}+\cdots 
\end{equation}
to expand the inverse operator in this equation, we can get the perturbation expansion of the effective Hamiltonian
\begin{eqnarray}\label{eq:EffectiveHamiltonian}
H_{\mathrm{eff}}(\lambda) & = & H_{ PP}+ V_{PP\perp}\frac1{\lambda -H_0^\perp} V_{P\perp P}
\nonumber\\
& + &  V_{PP\perp}     \frac1{\lambda -H_0^\perp}V_{P\perp P\perp}\frac1{\lambda -H_0^\perp}   V_{P\perp P}
\nonumber\\
&+& \cdots,
\end{eqnarray}
where $H_0^\perp = P^\perp H_0 P^\perp$.

\section{1st-Order Results: Transition Probability for Different Hamiltonians}
In terms of Eq.~(\ref{eq:FloquetHamiltonian}) and Eq.~(\ref{eq:ProjectionOperator}),we retain the effective Hamiltonian to the first-order perturbation term
\begin{equation}
H_{\mathrm{eff}}(\lambda) \approx
H_{ PP} = 
\begin{pmatrix}
E_\alpha & \langle \alpha | H^1 | \beta \rangle \\
\langle \alpha | H^1 | \beta \rangle^* & E_\beta + \omega
\end{pmatrix},
\end{equation}
where $\langle \alpha | H^1 | \beta \rangle$ is the upper triangular element, determined by the first positive Fourier component of the periodic Hamiltonian in Eq.~(\ref{eq:periodicHamiltonian}).
For periodic Hamiltonian $\mathcal H_1(t), \mathcal H_2(t), \mathcal H_4(t)$, the off-diagonal element of the effective Hamiltonian is the same
\begin{equation}
\langle \alpha | H^1 | \beta \rangle =b.
\end{equation}
Eigenvalues and eigenvectors of the effective Hamiltonian are
\begin{equation}\begin{aligned}
\lambda_\pm &= \frac 1 2 (\omega \pm \sqrt{4b^2 + (\omega_0 - \omega)^2 }),
\\
P | \lambda_\pm \rangle &= C_\pm
\begin{pmatrix}
\frac{ \omega_0-\omega \pm \sqrt{4b^2 + (\omega_0 - \omega)^2 } }{2b} \\
1
\end{pmatrix},
\end{aligned}\end{equation}
where $C_\pm$ is the normalization factor.
In fact, the energy gap between the two energy levels in a degenerate subspace is close to 0, while the energy levels outside the degenerate subspace have a finite energy gap with the energy levels inside the subspace.
This means that other undisturbed states have little impact on the dynamics inside the degenerate subspace, and $| \lambda_\pm \rangle \approx P | \lambda_\pm \rangle$.
From this, we can obtain two terms in Eq.~(\ref{eq:TimeEvolutionOperator}), corresponding to $\gamma l = \alpha 0,\beta 1$.
For other terms, we can similarly partition degenerate subspaces to obtain corresponding eigenstates, but their components $\langle\beta 1|\lambda_{\gamma l}\rangle,  \langle\lambda_{\gamma l}|\alpha0\rangle$ outside the subspace are very small, so we can ignore them directly in Eq.~(\ref{eq:TimeEvolutionOperator}).
It is easy to derive that
\begin{equation}\begin{aligned}
U_{\beta\alpha}(t;0)&\approx \langle\beta1|\lambda_{+}\rangle  \langle\lambda_{+}|\alpha0\rangle
\exp[-i\lambda_{+}t] e^{i\omega t}
\\
& + \langle\beta1|\lambda_{-}\rangle  \langle\lambda_{-}|\alpha0\rangle
\exp[-i\lambda_{-}t] e^{i\omega t}
\end{aligned}\end{equation}

Substituting it into Eq.~(\ref{eq:TransitionProbability}), we can obtain
\begin{equation}
P_{\alpha \to \beta}(t;0) = \frac{4b^{2}}{4b^{2}+\delta^{2}}\sin^{2}\frac{1}{2}\sqrt{4b^{2}+\delta^{2}}t,
\end{equation}
where $\delta = \omega - \omega_0$.
From this, the resonance condition $\omega=\omega_0$ is obtained, and the transition probability takes its maximum at $\delta=0$.
The corresponding oscillation is known as the Rabi oscillation, whose frequency is determined by the driving EM field's strength $b$ and the detuning $\delta$, giving the well-known generalized Rabi frequency, $\Omega_R=\sqrt{4b^2+\delta^2}$.
When resonance occurs, particles constantly undergo transitions between two spin states, emitting or absorbing photons.

For Hamiltonian $\mathcal H_3(t)$, the off-diagonal element of the effective Hamiltonian is
\begin{equation}
\langle \alpha | H^1 | \beta \rangle =0.
\end{equation}
It leads to the transition probability
\begin{equation}
P_{\alpha \to \beta}(t;0) = 0.
\end{equation}

Considering the quantum states of the total atomic and photon system, the states $|\alpha, N\rangle$ and $|\beta, N+1\rangle$ are degenerate and share the same energy.
The transition from $|\alpha, N\rangle$ to $|\beta, N+1\rangle$ represents that a high-energy atom emitting a photon.
Our results indicate that this transition does not always occur.
For the three electromagnetic wave modes corresponding to the Hamiltonian $\mathcal H_1(t), \mathcal H_2(t), \mathcal H_4(t)$, the off-diagonal element $\langle \alpha | H^1 | \beta \rangle = b$, meaning that there exists an interaction between quantum states $|\alpha, N\rangle$ and $|\beta, N+1\rangle$, leading to the occurrence of transitions.
For the electromagnetic wave mode corresponding to the Hamiltonian $\mathcal H_3(t)$, the interaction between two quantum states is $0$, which means that the two states are not related to each other.
Under this situation, a transition will not occur.
In the language of RWA, this spin resonance mode corresponds to the counter-rotating term with a negligible transition probability.
However, this result is given as a straightforward outcome in a framework combining the Floquet theory and Brillouin-Wigner perturbation method, proving the benefit of our new method dealing with the spin resonance.

\begin{figure}[h!]
\centering
\includegraphics{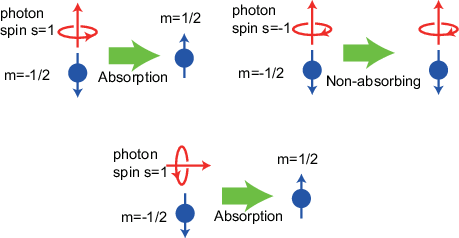}
\caption{Possible diagrammatic representation of resonance or absence of resonance caused by different types of photons.}
\label{fig2}
\end{figure}

For resonance corresponding to the Hamiltonian $\mathcal H_1(t)$, linear polarized photons can be seen as the superposition of left-handed circularly polarized $\sigma^-$ photons and right-handed circularly polarized $\sigma^+$ photons, exhibiting spin angular momentum parallel or antiparallel to the propagation direction \cite{Saleh2019}.
When a particle transitions from the $m=-1/2$ state to the $m=1/2$ state, it absorbs a $\sigma^+$ photon to increase the angular momentum along the $z$-axis by $\hbar$.
In this process, angular momentum along $z$-axis is conserved.
For resonance corresponding to the Hamiltonian $\mathcal H_2(t)$, only $\sigma^+$ photons exist.
The particle similarly absorbs a photon to increase the angular momentum along the $z$-axis.
For resonance corresponding to the Hamiltonian $\mathcal H_4(t)$, the projection of photon spin vector on the $y$-axis is $\hbar$.
Considering the conservation of angular momentum along the $y$-axis, photons may exert a torque on a particle, or photons that are related to the static magnetic field may participate in resonance processes.
The diagrammatic representation of transition is illustrated in Fig. \ref{fig2}.

For monochromatic electromagnetic waves propagating in any direction, we can usually decompose them into two parts: parallel and perpendicular to the static magnetic field. The Hamiltonian can still be expressed the same as Eq.\ref{eq:periodicHamiltonian}, the poposed method here is still effective.

\section{2nd-Order Results: Bloch-Siegert Shift}
For $\mathcal H_1(t)$, we retain the effective Hamiltonian to the second-order perturbation term and can obtain
\begin{eqnarray}
H_{\mathrm{eff}}(\lambda)&\approx&
H_{ PP}+ V_{PP\perp}\frac1{\lambda -H_0^\perp} V_{P\perp P}
\nonumber \\
&=&
\begin{pmatrix}
E_\alpha + \frac{b^2}{\lambda  -E_\beta + \omega} & b\\
b & E_\beta + \omega + \frac{b^2}{\lambda  -E_\alpha -2 \omega}
\end{pmatrix}
\end{eqnarray}
We refer to the idea in  Ref.~\onlinecite{Ho1984} and replace $\lambda$ with 0-order energy $E_\alpha$ and $E_\beta+\omega$.
We define
\begin{equation}
\Delta E = \lambda - \frac 1 2 ( E_\alpha + E_\beta + \omega ) = \lambda-\frac \omega 2,
\end{equation}
and get
\begin{equation}\begin{aligned}
\frac{1}{\lambda-E_{\beta}+\omega}&=\frac{1}{\frac{3}{2}\omega-E_{\beta}} \left(1 + \frac{\Delta E}{\lambda-E_{\beta}+\omega} \right),
\\
\frac{1}{\lambda-E_{\alpha}-2\omega}&=\frac{1}{-\frac{3}{2}\omega-E_{\alpha}} \left(1 + \frac{\Delta E}{\lambda-E_{\alpha}-2\omega} \right).
\end{aligned}\end{equation}
The second term on the right side of the equation is higher order perturbation. We keep the Hamiltonian only up to the second order
\begin{equation}\begin{aligned}
H_{\mathrm{eff}}\approx
\begin{pmatrix}
E_\alpha + \frac{b^2}{\frac 3 2 \omega + \frac 1 2 \omega_0} & b\\
b & E_\beta + \omega - \frac{b^2}{\frac 3 2 \omega + \frac 1 2 \omega_0}
\end{pmatrix}.
\end{aligned}\end{equation}

By calculating eigenstates $\lambda_\pm$ and eigenvalues $|\lambda_\pm \rangle$ of the effective Hamiltonians and substituting them into the expression of the time evolution operator, we can get the corresponding transition probability
\begin{equation}\begin{aligned}
P_{\alpha \to \beta} & (t;0) = | U_{\beta\alpha}(t;0) |^2 = \frac{4 b^2}{\left(\frac{4 b^2}{\omega_0+3 \omega }+\omega_0-\omega \right)^2+4 b^2}
\\
&\times
\sin ^2 \left[t \sqrt{\left(\frac{4 b^2}{\omega_0+3 \omega }+\omega_0-\omega \right)^2+4 b^2}\right]
\end{aligned}\end{equation}
The resonance condition is
\begin{equation}
\begin{aligned}
\omega &= \frac{1}{3} \left(2 \sqrt{3 b^2+\omega_0^2}+\omega_0\right)
\\
& \approx \omega_0  + (\frac{b}{\omega_0})^2 \omega_0.
\end{aligned}
\end{equation}
Compared to the result of first-order perturbation, the resonance condition has an additional frequency shift.
Such a shift of resonance frequency is called Bloch-Siegert shift \cite{Bloch1940}.
Our result agrees with previous ones \cite{Yan2015, Shirley1965, Beijersbergen1992}.

\section{Conclusions}
In this work, we studied the two-level spin resonance under four different electromagnetic wave modes based on the Floquet theory and perturbation method. Based on the first-order perturbation results, we have explained the reason why some electromagnetic waves can cause resonance while others cannot. The upper triangular element $\langle \alpha | H^1 | \beta \rangle$, determined by the first positive Fourier component of the periodic Hamiltonian, plays a decisive role. Our second-order perturbation result provides the Bloch-Siegert shift.
The method is analytical and intuitively understandable. It provides a new theoretical tool for investigating spin magnetic resonance, potentially useful in dealing with various quantum periodic pertuabtion problems.


\begin{acknowledgments}
G. Liu thanks the finiancial support from the Chinese Academy of Sciences (Grant No. E209YC1101).
\end{acknowledgments}


%

\end{document}